\documentclass[]{spie}  
\pdfoutput=1

\usepackage{amsmath,amsfonts,amssymb}
\usepackage{graphicx}
\usepackage[colorlinks=true, allcolors=blue]{hyperref}

\newcommand\dslm{Dragonfly Spectral Line Mapper}

\title{The Dragonfly Spectral Line Mapper: Design and First Light}

\author[a,b]{Seery Chen}
\author[c]{Deborah M. Lokhorst}
\author[a,b]{Jeff Shen}
\author[d]{Imad Pasha}
\author[e]{Evegni I. Malakhov}
\author[a,b]{Roberto G. Abraham}
\author[d]{Pieter van Dokkum}

\affil[a]{David A. Dunlap Department of Astronomy \& Astrophysics,
University of Toronto,
50 St. George Street, 
Toronto, ON M5S3H4, Canada}
\affil[b]{Dunlap Institute,
University of Toronto,
50 St. George Street, 
Toronto, ON M5S3H4, Canada}
\affil[c]{NRC Herzberg Astronomy \& Astrophysics Research Centre,
5071 West Saanich Road, 
Victoria, BC V9E2E7, Canada}
\affil[d]{Department of Astronomy,
Yale University,
52 Hillhouse Ave., New Haven, CT 06511, USA}
\affil[e]{New Mexico Skies, Inc., 
9 Contentment Crest, Mayhill, NM 88339, USA}

\authorinfo{Send correspondence to S.C\\ S.C: E-mail: schen@astro.utoronto.ca}

\begin{document} 
\maketitle

\begin{abstract}
The Dragonfly Spectral Line Mapper (DSLM) is the latest evolution of the Dragonfly Telephoto Array, which turns it into the world's most powerful wide-field spectral line imager. The DSLM will be the equivalent of a 1.6m aperture $f$/0.26 refractor with a built-in Integral Field Spectrometer, covering a five square degree field of view. The new telescope is designed to carry out ultra-narrow bandpass imaging of the low surface brightness universe with exquisite control over systematic errors, including real-time calibration of atmospheric variations in airglow. The key to Dragonfly's transformation is the ``Filter-Tilter", a mechanical assembly which holds ultra-narrow bandpass interference filters in front of each lens in the array and tilts them to smoothly shift their central wavelength. Here we describe our development process based on rapid prototyping, iterative design, and mass production. This process has resulted in numerous improvements to the design of the DSLM from the initial pathfinder instrument, including changes to narrower bandpass filters and the addition of a suite of calibration filters for continuum light subtraction and sky line monitoring. Improvements have also been made to the electronics and hardware of the array, which improve tilting accuracy, rigidity and light baffling. Here we present laboratory and on-sky measurements from the deployment of the first bank of lenses in May 2022, and a progress report on the completion of the full array in early 2023.
\end{abstract}

\keywords{Low surface brightness imaging, narrowband interference filters}

\section{INTRODUCTION}
\label{sec:intro}  

One of the ultimate known unknowns in the field of extragalactic astronomy is the circumgalactic medium (CGM). The CGM forms an enormous reservoir of diffuse gas surrounding each galaxy, extending far beyond the central region of galaxies where stars exist.  Understanding how gas flows into and out of the CGM and the intergalactic medium (IGM) through processes such as recycling and feedback, and how this in turn influences star formation, is crucial for understanding galaxy evolution. However, the CGM is poorly constrained by observation. 
Conventional observations of the CGM/IGM can be categorized into roughly three approaches: radio observations of the neutral component \cite{2007_Oosterloo_radioNCG891, 2011_heald_21cm_halogas, 2017_moss_21cm_densediffuse, 2018_emonts_molec_radio}, X-ray observations of hot gaseous coronae \cite{2011_anderson_xray, 2012_Dari_xray, 2013_bogdan_xray, 2015_walker_xray, 2020_das_xray}, and absorption line studies which probe various phases of the CGM/IGM \cite{2013_werk_coshalos_cgm, 2016_danforth_coshalos_igm, 2017_richter_coshaloslegacy_cgm ,2011_tumlinson_0VI}. The majority of observational constraints for the CGM/IGM have been yielded by absorption line studies that look at the spectra of background quasars. However, a major downside to these absorption line studies is they can only probe the CGM indirectly along narrow “pencil beam” lines of sight, and as such are limited in their ability to investigate the spatial morphology of the CGM structure.  An alternative approach to investigate the CGM/IGM that has recently become feasible through technological upgrades to telescope instrumentation is direct imaging of the IGM and CGM via UV/visible emission lines. At $T \sim 10^5$K the CGM and IGM cools radiatively, giving off very weak emission lines, with Ly$\alpha$ $\lambda1216$ being the strongest of these \cite{2010FaucherLyalpha}. The sensitivity required to detect this diffuse Ly$\alpha$ emission was unreachable until the advent of spatially resolved spectrometers such as the Keck Cosmic Web Imager (KCWI) \cite{keck2018-CWI} at the W. M. Keck Observatory and the Multi Unit Spectroscopic Explorer (MUSE) \cite{muse2010-instrument} at ESO's Very Large Telescope (VLT). These integral field spectrometers have allowed this field of study at high redshifts to progress rapidly\cite{wisotzki2016-musehalos, Leclerq2020-musehalos,Daddi2021-filaments, muse2021-cosmicweb}, however little progress has been made at low redshifts. This is due to a combination of the inaccessibility of the Ly$\alpha$ line to ground-based studies at redshifts $z<2.5$ and the limited fields of view ($<$1 arcmin$^2$) of these integral field spectrometers, which are imperfectly suited to nearby targets at cosmological distances of D $\lesssim$ 40 Mpc. 
Our solution is to directly image the CGM of local galaxies in H$\alpha$ emission. This emission is very faint, requiring an estimated lower surface brightness detection limit of $7\times10^{-20}$ erg  s$^{-1 }$  cm$^{-2}$ arcsec$^{-2}$ (1000 ph s$^{-1 }$ cm$^{-2}$ sr$^{-1}$) \cite{lokhorst2019, 2011steidel}, deeper than the deepest low surface brightness H$\alpha$ wide-field imaging available in the literature, which reach depths of $10^{-19}$ to $10^{-18}$ erg  s$^{-1 }$  cm$^{-2}$ arcsec$^{-2}$.  \cite{1995leo,halphatail-limit,m101-lowSB2017}. Thus, to image the CGM/IGM directly at low redshift requires a telescope capable of unprecedentedly low surface brightness emission line imaging.

Low surface brightness imaging is where the Dragonfly Telephoto Array (Dragonfly) excels. Dragonfly is a mosaic refractor telescope comprised of an array of commercial high-end Canon telephoto lenses (400 mm f/2.8 Canon IS II) \cite{2014Dragonfly}. In its current 48-lens configuration, Dragonfly is equivalent to a 1 m aperture, $f$/0.4 telescope, with a  $2.6^{\circ}\times1.9^{\circ}$ field of view (FOV). The limiting factors for low surface brightness imaging are systematics such as scattered light contamination and calibration accuracy, as opposed to photon noise (which is often the limiting factor for point sources, and be reduced with bigger telescopes \cite{2017galbook}). Dragonfly reduces these systematics in multiple ways. For example, by being an all lens (refractor) telescope, Dragonfly avoids many issues with scattered light that arise from large angle scattering from reflective surfaces. The lack of a central obstruction in the optical path of the lenses helps keep the wings of the point spread function well controlled. Lenses also avoid the backscatter of light into the optical path caused by dust and microroughness on reflective coatings found in reflector systems. The Canon telephoto lenses used by Dragonfly are well baffled and have a nanostructure-based anti-reflection coating that minimizes ghosting and flare caused by internal reflections.

The Dragonfly design yields incredibly optically ``fast" (low focal ratio) telescopes, making it well suited for low surface brightness imaging. This is because the imaging speed needed to achieve a fixed signal-to-noise ratio for very extended structures (much larger than the resolution limit) depends on the focal ratio, not the aperture. By pointing all Dragonfly lenses at the same position and stacking the frames obtained, the effective diameter of the telescope scales as the square root of the number of lenses, and the effective focal ratio scales as the inverse. Additionally, Dragonfly's wide FOV allows for better sky subtraction that is crucial for measuring the extent and luminosity of low  surface  brightness  objects. These aspects combined make Dragonfly uniquely well suited to imaging spatially very extended, extremely low surface brightness structures. 

The challenge of introducing narrowband interference filters to a fast optical system is that the performance of such filters degrades in fast light cones and varies as a function of field angle. These filters rely on constructive and destructive interference of the wave-front,  so their efficiency depends on the optical path length through the coating stack. This, rather than the ability to manufacture narrow bandpass filters, has been the limiting factor on bandwidth for narrow band imaging on conventional wide field telescopes. Our solution to this ``problem" turns  the field angle dependence in interference filters into a ``feature".  Individual Dragonfly lenses are small enough (143mm aperture) that full-aperture interference filters can be positioned in front of each lens. By placing the filter at the entrance pupil of the optical system, the incoming beam is non-converging (with an $f$-ratio of infinity). This allows Dragonfly to observe with a much narrower bandwidth than traditional telescopes which place interference filters in a fast converging beam \cite{lokhorst2020}.  Tilting these filters shifts their central wavelength, allowing Dragonfly to tune to the H$\alpha$ line at the redshift of the target galaxy, with a range encompassing thousands of galaxies within the local Universe.

We have obtained funding for a 120-lens Dragonfly array fitted with specialty narrow band filters. This telescope will be capable of directly imaging down to a surface brightness of 1000 ph s$^{-1 }$ cm$^{-2}$ sr$^{-1}$ \cite{lokhorst2019}. Rather than outfit the existing Dragonfly Telephoto Array with ultra-narrow bandpass filters, we are constructing a new array using the Dragonfly design as a basis. We refer to this new array with ultra-narrow bandpass filters the Dragonfly Spectral Line Mapper. A 3-lens pathfinder instrument with ultra-narrow bandpass H$\alpha$ filters was on sky from March 2020 to October 2021, providing a proof of concept \cite{lokhorst2020} and obtained scientifically significant observations \cite{lokhorst2022shell,pasha2021}. Construction of the 120-lens Dragonfly Spectral Line Mapper has now started, with 10 of the 120 lenses on sky and taking data as of May 2022. This paper will describe the design of the Dragonfly Spectral Line Mapper, it's current 10-lens array configuration, the testing of various components of the array, and a progress report on the completion of the full 120-lens array.

This paper is one of three in a series in this SPIE Proceedings Volume on imaging the low surface brightness universe with mosaic telescopes and the Dragonfly Telephoto Array. Table \ref{tab:3papers} summarizes the topics covered by each paper.

\begin{table}[ht]
\caption{Content summary of the three papers in this series appearing in this SPIE Proceedings, with the topics covered by this paper in bold.} 
\label{tab:3papers}
\begin{center}       
\begin{tabular}{|l|l|} 
\hline
\rule[-1ex]{0pt}{3.5ex}  Subject & Paper in this Volume \\
\hline\hline
\rule[-1ex]{0pt}{3.5ex}  Mosaic telescope general concepts & Abraham et al. \\
\hline
\rule[-1ex]{0pt}{3.5ex}  Mosaic telescope design trades	 & Abraham et al. \\
\hline
\rule[-1ex]{0pt}{3.5ex}  Low surface brightness imaging challenges & Abraham et al.  \\
\hline
\rule[-1ex]{0pt}{3.5ex}  Narrowband imaging concepts and methods & Lokhorst et al.  \\
\hline
\rule[-1ex]{0pt}{3.5ex}  Narrowband imaging survey speed & Lokhorst et al. \\
\hline
\rule[-1ex]{0pt}{3.5ex}  Dragonfly Spectral Line Mapper pathfinder results \& lessons learned & Lokhorst et al. \\
\hline
\rule[-1ex]{0pt}{3.5ex}  \textbf{Dragonfly Spectral Line Mapper design } & Chen et al. \\
\hline
\rule[-1ex]{0pt}{3.5ex}  \textbf{Dragonfly Spectral Line Mapper laboratory tests}	 & Chen et al.\\
\hline
\rule[-1ex]{0pt}{3.5ex}  \textbf{Dragonfly Spectral Line Mapper roadmap} & Chen et al. \\
\hline
\end{tabular}
\end{center}
\end{table}

\section{Instrument Overview}\label{sec:design}

\begin{figure}[h]
	\centering
    \includegraphics[width=1\textwidth]{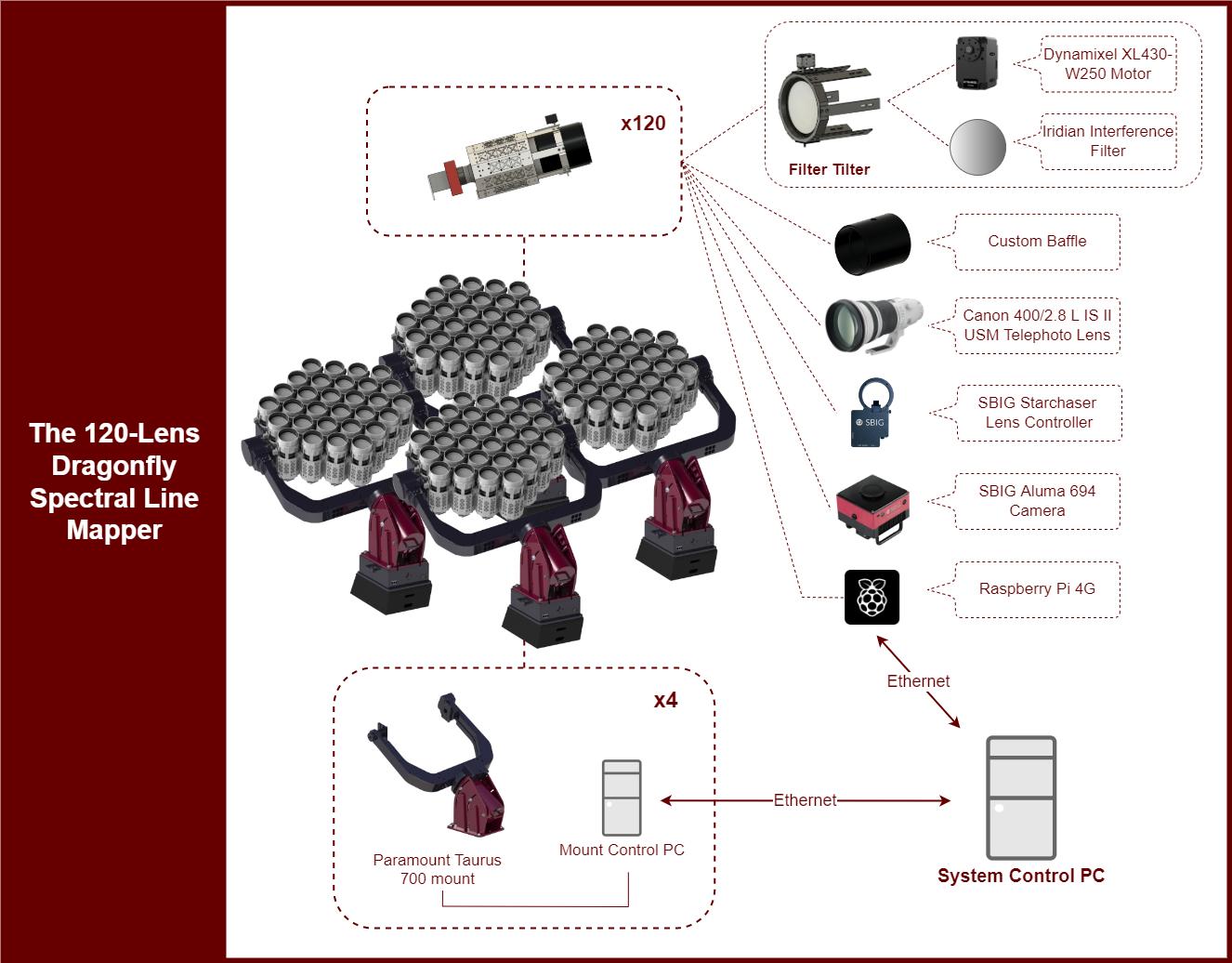}
    \caption{Schematic view of the proposed 120-lens setup of the Dragonfly Spectral Line Mapper. Each mount holds 30 lenses and will be controlled with its own mount control PC. Each lens is a part of a fully independent subsystem with its own Filter-Tilter, baffle, lens controller (focusing and and image stabilization), camera, and Raspberry Pi. The Dragonfly Spectral Line Mapper and the original Dragonfly Telephoto Arrays are located at New Mexico Skies Observatory. Observing with these telescopes occurs remotely.}  
    \label{fig:schem}
\end{figure}

Figure \ref{fig:schem} shows a schematic view of the full 120-lens Dragonfly Spectral Line Mapper. In its final form, the Dragonfly Spectral Line Mapper will have four mounts, each holding an array of 30 lens units for a total of 120 lenses. Each lens unit is a fully independent subsystem, with all lens units controlled by a single control PC via an Internet of Things protocol. All 120 lenses are aligned to point to the same target creating the equivalent to a 1.6m, $f$/0.26 refractor telescope. Table \ref{tab:dslm} shows a summary of the specifications for reference.

The Filter-Tilter holds the filter in front of each lens, and rotates the filter to a specified angle with a servo actuator attached along the central axis of the filter. The filter is placed within the inner ring of the Filter-Tilter, and the inner ring is attached to the outer ring on two shafts connected to ball bearings. The actuator is attached to the end of one shaft. 

The updated baffle is a custom 3D printed design made of ABS plastic and coated with anti-reflective material (e.g., rayon flocking with airbrushed ultra-flat matte black paint). The baffle comprises three sections, the baffle front, baffle back, and the lens collar. The baffle front and baffle back attach the the outer ring of the Filter-Tilter and have curved inner walls that minimize the gap between the filter edges and the inner surface of the baffle, especially when the filter is tilted to higher tilt angles. The lens collar is secured to the baffle back and the lens, which makes the system more rigid. On the lens collar, the circular opening where the lens sits is offset by 2mm from the center, which allows for the lenses to be offset by up to +/- 0.28$^{\circ}$ from each other. When randomly orienting this offset, the effective FOV of all 120 lenses together becomes 2.5$^{\circ}\times$2.1$^{\circ}$.

The lens controllers control the internal auto-focusing motors and image stabilization motors within each lens. These lens controllers also serve as the physical connector between the camera and the lens. On the pathfinder as well as the current 10-lens array, we use custom units made by Birger Engineering which only have focusing capabilities. The full 120 lens array will use custom Starchaser lens controllers made by Diffraction limited, with the ability to send focus and image stabilization commands to the lens. The Starchasers also have an off-axis guider camera which will be used to inform the focusing and image stabilization of each individual lens. The use of active image stabilization on the Dragonfly Spectral Line Mapper is currently in development and will be discussed in a future paper.

The cameras used are the SBIG Aluma CCD694 cameras manufactured by Diffraction Limited, the same as the cameras used for the pathfinder. These cameras use a Sony ICX-694 CCD detector, with a pixel scale of 2.5" and a pixel size of 4.54 $\mu$m. The detector is 2200x2750 pixels, corresponding to a FOV of 1.5$^{\circ}\times$1.9$^{\circ}$. The quantum efficiency ranges between 70-75\% between 450-620nm and drops to 40\% near 750nm. The specified read noise is 4.5 electrons (RMS), and the dark current is 0.025 electron pixel$^{-1}$ s$^{-1}$. In our testing we find the read noise to be closer to 4 electrons, and the dark current to be an order of magnitude lower than specified at 0.0026 electrons pixel$^{-1}$ s$^{-1}$ in the center of the frame and increasing to to 0.025 electrons $^{-1}$ s$^{-1}$ only at the edge of the frame in the presence of amplifier (amp) glow. Further details are presented in Section \ref{sec:aluma}.

The 120-lens array will use the Paramount Taurus 700 mounts manufactured by Software Bisque Inc. This is a direct drive equatorial fork mount with 270kg total instrument capacity, one of the largest commercial off-the-shelf (COTS) mounts available. Using four COTS mounts in four COTS domes rather than a custom large mount and dome greatly reduces the total cost of our facility. For testing the 10-lens array, we are currently using the Paramount ME II mount manufactured by Software Bisque Inc.

\begin{table}[ht]
\caption{Dragonfly Spectral Line Mapper specifications for the 120-lens array.} 
\label{tab:dslm}
\begin{center}       
\begin{tabular}{|l|l|} 

\hline
\rule[-1ex]{0pt}{3.5ex}  \textbf{Parameter} & \textbf{Value} \\
\hline\hline
\rule[-1ex]{0pt}{3.5ex}  Optics & 120X Canon Telephoto 400mm $f$/2.8 IS II \\
\hline
\rule[-1ex]{0pt}{3.5ex}  Camera & 120X Diffraction Limited Aluma 694 \\
\hline
\rule[-1ex]{0pt}{3.5ex}  Filters & See Table \ref{tab:DSLMFilters}   \\
\hline
\rule[-1ex]{0pt}{3.5ex}  Mount & 4X Paramount Tarus 700   \\
\hline
\rule[-1ex]{0pt}{3.5ex}  Sensor & Sony ICX 694 CCD, one per camera \\
\hline
\rule[-1ex]{0pt}{3.5ex}  Effective diameter (m) & 1.6 \\
\hline
\rule[-1ex]{0pt}{3.5ex}  Effective focal ratio & 0.26 \\
\hline
\rule[-1ex]{0pt}{3.5ex}  Effective FOV & 2.1$^\circ \times 2.5^\circ$  \\
\hline
\rule[-1ex]{0pt}{3.5ex}  Location & New Mexico Skies Observatory \\
\hline
\end{tabular}
\end{center}
\end{table}


\section{Filter Specifications}\label{sec:filterspecs}

\subsection{Emission line filters}

The goal of the Dragonfly Spectral Line Mapper instrument is to survey the CGM and IGM in the local Universe. A crucial decision for the upgrade is the central wavelength (CWLs) of the narrowband filters, since the CWL determines not only the line transition(s) that can be targeted, but also the volume of Universe that can be probed with the filter. The Filter-Tilter design allows a range of wavelengths larger than the bandpass of the narrowband filter to be imaged. As shown in Ref.~\citenum{lokhorst2020}, the CWL of an interference filter is shifted to shorter wavelengths following a well-defined relationship with the angle of incidence of light upon the filter, making available an additional wavelength range of $\sim$8 nm. It is important to note that the intrinsic CWL (i.e., the CWL at zero-angle of incidence) is at the long end of that range $-$ tilting the filters to larger angles always shifts the CWL to shorter values. 
The choice of the CWL is therefore a critical design element in the instrument. In addition to the ultranarrowband filters, off-band filters are also required to calibrate and subtract background contamination from the data. The chosen filter specifications are listed in Table~\ref{tab:DSLMFilters} and the reasoning for these specifications is given below.

The emission lines chosen to be targeted with the \dslm~are H$\alpha~\lambda6563$, \textsc{[Nii]}~$\lambda6583$ and \textsc{[Oiii]}~$\lambda5007$. In Ref.~\citenum{lokhorst2019}, we demonstrated that emission from the CGM radiated by both the H$\alpha$ and \textsc{[Oiii]} emission lines would be visible with suitably long integrations. The \textsc{[Nii]} line is much fainter, but its proximity to the H$\alpha$ wavelength will allow us to target both lines with one filter, so only two different types of narrowband filter will be needed.
The maximum redshift was chosen based on the radial velocity distribution of the Virgo Galaxy Cluster.
The radial velocities of galaxies in the Virgo Cluster range from $\approx 500 - 2900$ km/s. 
In order to image emission from galaxies with this full range of recessional velocities, the filter central wavelengths must be at the upper end of the range, which corresponds to a redshift of $z \sim 0.01$.
This redshift shifts the H$\alpha$/\textsc{[Nii]} line from its rest wavelength of $\lambda$ = 656.3/658.3 nm to 662.6/664.7 nm\footnote{Note that all wavelengths used in this work are in air.}. As the CWL can only be shifted to shorter values, the higher number, 664.7 nm, is chosen for the CWL of the H$\alpha$ filter. The shift in the CWL as a function of tilt is modelled and shown in section \ref{sec:filters}.
With a CWL of 664.7 nm, the effective CWL is shifted to the rest wavelength of H$\alpha$ at a tilt of about 19$^{\circ}$, so the whole of the local volume can be probed with a tilt range of 20$^{\circ}$. The aperture diameter of the filter has been chosen so that at this tilt the full aperture of the lenses is still being utilized. 
The same filter can track [NII] line emission, reaching the rest wavelength of [NII] at a tilt of about 16$^{\circ}$.
The properties of the \textsc{[Oiii]} filters were defined to match the redshift range able to be mapped in H$\alpha$ (i.e., a recessional velocity range of v $\approx 0 - 3900$ km/s). With an intrinsic CWL of 507.1 nm, the rest wavelength of \textsc{[Oiii]}, $\lambda = 500.7$ nm, is reached at a tilt of $\approx 19 ^{\circ}$.
As the maximum tilt that the filters can be rotated to is $\approx 20^{\circ}$, the chosen CWLs will enable imaging of all three emission lines at recessional velocities from $\approx 0 - 2900$ km/s. This range of $20^{\circ}$ is a fundamental design limit of the Filter-Tilters.

 \begin{table}[ht]
\caption{Filters specifications for the Dragonfly Spectral Line Mapper. All Filters are 154.4mm in diameter with a clear aperture $>$150nm, and a surface roughness $<$2.0nm. All filters are $>$OD4 blocking over 200-1000nm excluding the band pass region (approximately 10 nm from bandpass edge), with the OH Off filter blocking OD4 at 771nm as well. } 
\label{tab:DSLMFilters}
\begin{center}       
\begin{tabular}{|l|l|l|l|}
\hline
\rule[-1ex]{0pt}{3.5ex} Filter  &	Central Wavelength & Bandpass & Number on Array \\
                                &     	 (nm)          &  (nm)    &     \\
\hline\hline
\rule[-1ex]{0pt}{3.5ex} H$\alpha$                       &	    664.7          &   0.8    &  60 \\
\hline
\rule[-1ex]{0pt}{3.5ex} H$\alpha$ Off $-$ Right         &	    611.0          &   30     &  3  \\
\hline
\rule[-1ex]{0pt}{3.5ex} H$\alpha$ Off $-$ Left          &	    705.0          &   30     &  3  \\
\hline
\rule[-1ex]{0pt}{3.5ex} \textsc{[Oiii]}                 &	    507.1          &   0.8    &  40 \\
\hline
\rule[-1ex]{0pt}{3.5ex} \textsc{[Oiii]} Off $-$ Right   &	    468.0          &   30     &  3  \\
\hline
\rule[-1ex]{0pt}{3.5ex} \textsc{[Oiii]} Off $-$ Left    &	    538.5          &   30     &  3  \\
\hline
\rule[-1ex]{0pt}{3.5ex} OH On                           &	    772.0          &   2      &  4 \\
\hline
\rule[-1ex]{0pt}{3.5ex} OH Off                          &	    768.75         &   2      &  4 \\

\hline 
\end{tabular}
\end{center}
\end{table}

\subsection{Calibration filters}

There are three background components that need to be subtracted from the data when carrying out narrowband or spectral analysis. These are the sky continuum, OH sky lines, and the continuum from astronomical sources. 
To improve the fidelity of background sky subtraction, our design provides for a small number of background calibration filters to be used to monitor the sky at the same time as the H$\alpha$ and \textsc{[Oiii]} filters are obtaining data on the science targets.

\subsubsection{Medium-band Filters}

Two sources of background continuum emission will be subtracted through the use of off-band filters which have wavelength ranges selected to avoid sky lines. These medium-band filters were chosen to have bandwidths in between the narrow- ($< 10$ nm) or broad-band ($> 100$ nm) ranges. The filters need to be narrow enough to avoid strong sky lines, yet large enough that they transmit a sufficiently large number of continuum photons without requiring an equivalent area or exposure time to the science filters. A proportion of about 30:1 for the number of medium-band to narrowband filters was chosen, which will require only one medium-band filter per mount (which will contain 30 filters total). For both science emission line filters, we choose to implement two medium-band filters with central wavelengths on either side of the emission line filter central wavelengths (H$\alpha$ Off $-$ Right/Left and \textsc{[Oiii]} Off $-$ Right/Left). This will allow any slope in the continuum over the wavelength region to be modeled out.

\subsubsection{OH Line Monitoring filters}

There may be a need for another calibration filter to subtract strong sky lines from the data. The OH lines in the sky spectrum are exceedingly bright, with increasing brightness and frequency at longer wavelengths. It is therefore most efficient to monitor an OH line at longer wavelengths where the OH line emission is the brightest, which would require a smaller number of OH line monitoring filters on the \dslm. A central wavelength of 772 nm was chosen for the OH line monitoring filter. This location has bright OH sky lines and the quantum efficiency of the CCD detector still remains high at this wavelength.

The sky continuum will be scaled out of the OH line monitoring filter using another filter with a CWL near that of the OH line monitoring filter and avoiding sky lines (a CWL of 768.75 nm was chosen for this filter). After sky continuum subtraction, the OH line data can be scaled and subtracted from the emission line data. The scaling will be based on the OH lines that fall within the bandpass of the emission line filter during the tilt chosen for each specific science exposure.
The ratio between the OH line emission at 772 nm and at lines near the H$\alpha$ wavelength can be calculated based on the molecular transitions \cite{oste96}. To first order, we can use groupings of OH lines with the same vibrational transitions to scale and subtract the sky background.  There are residuals from this, which require subtraction based on lines with the same OH rotation transitions to correct. If  rotational temperature varies, lines can be either over- or under-subtracted \cite{davi07} so careful testing of the filters will be required to create a robust methodology for subtracting atmospheric OH line emission from the science data.

\section{LABORATORY TESTING of Filters}\label{sec:filters}

Laboratory testing characterized the bandpass as a function of filter tilt angle for the 0.8nm H$\alpha$ and \textsc{[Oiii]} filters. To obtain spectra, we used the LHIRES III spectrometer at the end of a one lens unit of the Dragonfly Spectral Line Imager in a laboratory setting. We created an approximately collimated light source with a taped LED serving as a ``point source" and another Canon lens serving as a collimating lens, and pointed it at the one lens unit. The ``zeropoint" of the Filter-Tilter, where the surface normal vector of the filter is parallel with the optical axis, was found by self collimating the light source. The spectrometer was calibrated with a Neon lamp for the H$\alpha$ filter spectra, and a Neon-Argon lamp for \textsc{[Oiii]} filter spectra.

The Filter-Tilter tilted the filter in increments of two degrees from -20$^{\circ}$ to 20$^{\circ}$ and a spectra was taken at each tilt. The spectra were dark subtracted and divided by the spectra of the light source. The resulting transmission curves were well approximated by a second order ``super-Gaussian" (eq. \ref{supergaussian}), and thus were fit to eq. \ref{supergaussian} to obtain the central wavelength (CWL), 
\begin{equation}\label{supergaussian}
    T = A \exp{ (\frac{-(\lambda-\lambda')^4}{4\sigma^4}})
\end{equation}
where T is the transmission, A is a scaling factor, $\lambda$ is the wavelength, $\lambda'$ is the CWL of the bandpass, and $2(4 \log 2)^{1/4} \sigma$ is the FWHM of the bandpass. 

\begin{figure}[h]
	\centering
    \includegraphics[width=0.56\textwidth]{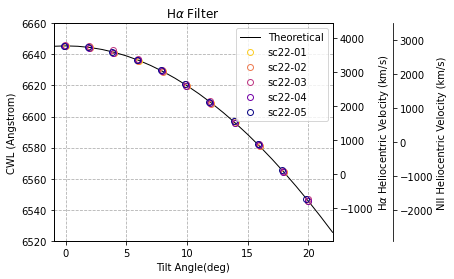}
    \caption{Central wavelength of the 0.8nm H$\alpha$ filter bandpass as a function of the angle of incidence to the filter (the ``tilt" angle of the Filter-Tilter). This angle is measured by the encoder within the DYNAMIXEL XL430-W250-T servo actuator. The spectrum at each tilt was fit to equation \ref{n2eq} to obtain the CWL. Five filters were tested and are overplotted here with the legend indicating the serial number of the filter. Lab measured CWL typically differ no more than 0.5 $\mathrm\AA$ from the theoretical value, where the effective index of refraction $n_{eff}=1.98$ and the central wavelength is 6647 $\mathrm\AA$.}  
    \label{fig:cwltilt3}
\end{figure}

The CWL $\lambda'$ at a given tilt angle $\theta$ is expected to follow
\begin{equation}\label{n2eq}
    \lambda' = \lambda_0 \sqrt{1- (\frac{1}{n_c}\sin{\theta})^2}
\end{equation}
with $\lambda_0$ being the rest wavelength of H$\alpha$ in air, and $n_c$ being the effective index of refraction of the filter. The fit to eq. \ref{n2eq} for five H$\alpha$ narrowband filters are shown figure \ref{fig:cwltilt3}. We find $n_c = 1.98$ for all H$\alpha$ filters. The shape of the transmission curve remains roughly the same from $0^{\circ}$ to $\pm20^{\circ}$ in filter tilt, varying by at most 40\% on the edges of the bandpass. Another way to quantify this is that the FWHM of the filter changes by from 8.5 $\mathrm\AA$ at zero tilt to 7.2 $\mathrm\AA$ at $\pm 20^{\circ}$ tilt. This slight change in bandpass shape is consistent across all five H$\alpha$ filters.

\begin{figure}[h]
	\centering
    \includegraphics[width=0.5\textwidth]{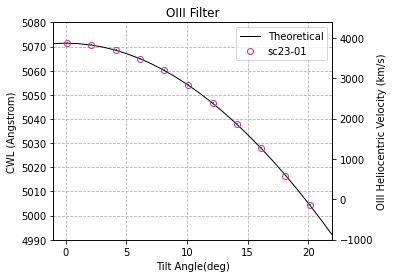}
    \caption{Central wavelength of the \textsc{[Oiii]} filter bandpass from 5071.3 $\mathrm\AA$ at zero tilt vs the tilt angle of the filter as measured by the DYNAMIXEL XL430-W250-T actuator. The spectrum at each tilt was fit to Equation \ref{n2eq} to obtain the CWL.}  
    \label{fig:oiicwltilt}
\end{figure}

Following the calculations done for the H$\alpha$ filters, we find $n_c = 2.12$ for the \textsc{[Oiii]} filter (see Figure \ref{fig:oiicwltilt}). The shape of the transmission curve remains roughly the same from $0^{\circ}$ to $\pm20^{\circ}$ in filter tilt, varying by at most 15\% at the edges of the bandpass. 

\section{Filter-Tilter ACCURACY}\label{sec:filtertilter}

The 0.8nm bandpass filters requires a position angle accuracy of 0.4$^{\circ}$ or better, which the original Filter-Tilter with a $\pm$ 1$^{\circ}$ accuracy did not meet \cite{lokhorst2020}. In the updated filter filter design we use a DYNAMIXEL XL430-W250-T servo actuator. This is a low cost, integrated servo system with a DC motor, gearbox, driver, and 12-bit on-axis magnetic rotary position sensor (AMS AS5601). Positioning is controlled using a Proportional Integral Derivative (PID) controller, using feedback from the 12-bit encoder to operate as a closed loop system. The gearbox is placed between the motor and the encoder to correct for any potential backlash \cite{dynamixel_train}. The small form factor of the DYNAMIXEL actuator allows it to sit on the central axis of the Filter-Tilter. This direct drive placement allows the angle sensor in the actuator to directly measure the rotation position angle of the filter. The AMS angle sensor used within the DYNAMIXEL actuator has a full scale linearity error of 0.11$^{\circ}$ in the optimal alignment of the magnet with the sensor chip (i.e over a full 360$^{\circ}$ of rotation, the error in the angle sensor measurement 0.11$^{\circ}$ at maximum). This linearity error increases to 1$^{\circ}$ at 1mm displacement. However, any non-linearity should be corrected for by the DYNAMIXEL actuators as they are calibrated by the manufacturer during assembly to correct for known positional error readings. 

\begin{figure}[h]
	\centering
    \includegraphics[width=0.85\textwidth]{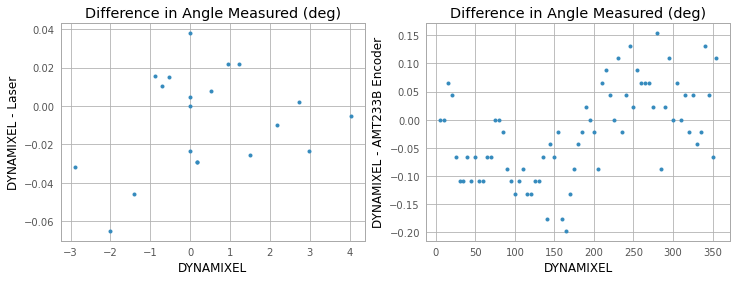}
    \caption{The difference between the position angle measure by the DYNAMIXEL actuator, laser method, and AMT233B-V encoder, with both x an y axis in degrees. }  
    \label{fig:motorlaser}
\end{figure}
 
The test setup for determining the accuracy of the DYNAMIXEL actuator is as follows: the DYNAMIXEL actuator was mounted on the Filter-Tilter, a mirror was placed in the filter frame, a laser pointer beam was positioned to reflect off of the mirror and onto the opposing wall 6m away. We tilted the Filter-Tilter in random increments going from -3 to 4$^{\circ}$. A larger range was not used due to the physical constraints of the room. At each tilt angle, we recorded the angle measure by the DYNAMIXEL actuator and measured the distance the laser point moved up and down the wall as the Filter-Tilter tilted to obtain a independent measurement of the ``true" tilt angle of the Filter-Tilter. In this controlled laboratory environment, we find the DYNAMIXEL actuator to be accurate within 0.1$^{\circ}$ of the laser angle measurement 100\% of the time, see figure \ref{fig:motorlaser}. 95\% of the time the DYNAMIXEL actuator was accurate to within 0.05$^{\circ}$. This shows the error is well within tolerance over small changes in position angle. 

As the linearity error of the AMS angle sensor follows a $\cos2x$ function (where $x$ is the position angle), the seven degree range is insufficient to conclude the size of linearity errors that are potentially uncorrected by the DYNAMIXEL actuator. We used the CUI Devices AMT233B-V angle sensor with an accuracy of 0.2$^{\circ}$ to compare with the actuator over a full 360$^{\circ}$ of rotation and find the angle reported from the DYNAMIXEL's angle sensor and the CUI AMT233B-V angle sensor to be within $\pm$0.2$^{\circ}$, and thus take 0.2$^{\circ}$ as the upper limit in the DYNAMIXEL actuator's angle measurement error, fulling the required specification of a tilt angle accuracy in the Filter-Tilter of 0.4$^{\circ}$ or better. With a higher accuracy encoder, one could determine if the DYNAMIXEL actuator has an accuracy higher than 0.2$^{\circ}$.

\section{ALUMA694 CAMERA CHARACTERIZATION}\label{sec:aluma}

To verify the gain of the Aluma 694 camera, we took a number of 30s exposure flat field images with a narrowband H$\alpha$ lens, equipped with a Flip-Flat (a luminescent flat field panel manufactured by Alnitak) set to varying illumination levels. The flats were then cropped around the center 200x200 pixels to minimize effects from vignetting, and the mean and variance was taken over the cropped section for each frame. As the photon counts follows a Poisson distribution, taking a linear fit of variance vs average number of counts, we get the gain as the inverse of the slope. This results in a gain of 0.24 electron ADU$^{-1}$.

\begin{figure}[h]
	\centering
    \includegraphics[width=0.9\textwidth]{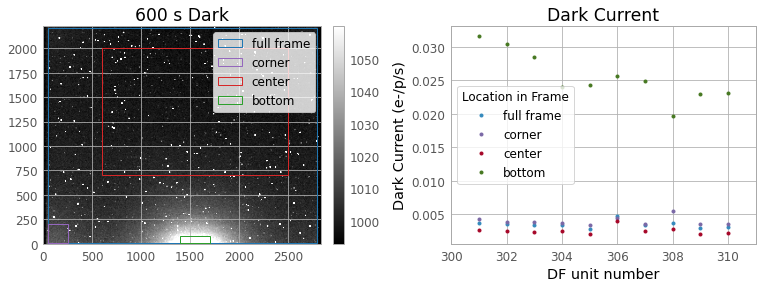}
    \caption{Left: 600s dark taken by the Aluma 697 camera, with coloured boxes overlaid to show locations in the frame where the dark current was calculated. Right: the dark current calculated for each cropped section of the frame. }  
    \label{fig:darkcurrent}
\end{figure}

To verify the read noise for one camera we cooled the detector to -20$^{\circ}$ C, took 10 bias frames and used the two bias subtraction method (two bias frames are subtracted from each other, and the read noise is the standard deviation of the subtracted frame divided by $\sqrt{2}$, multiplied by the gain). The read noise was calculated for every unique pairing of the 10 frames, and the final read noise for the camera was determined by taking the mean over all of them. Then this was repeated for all 10 cameras, and we found the read noise to be $4.0 \pm 0.3$ for the Aluma 694 camera, with the error indicating 1$\sigma$ variation in the read noise calculated between the 10 cameras.

To verify the dark current, we took 10 darks at each of the following exposure times: 5, 30, 600, 1800, and 3600s. As there was significant amp glow in some sections, we took the mean counts across a section of a dark with and without amp glow, and took a linear fit of counts vs exposure time, and multiplied by the gain. Figure \ref{fig:darkcurrent} shows where the darks were cropped for this calculation, and the resulting dark currents calculated. The center of the bottom edge of the frame on average has a dark current of 0.025 e-/p/s, approximately 10 times higher than the center of the frame with a dark current of 0.0026 e-/p/s.

\section{Development Roadmap}

The Dragonfly approach to instrumentation is to move fast and iterate with prototypes, so mistakes can be caught early. The 3-lens pathfinder instrument was on sky taking data from March 2020 to October 2021. The 10-lenses Dragonfly Spectral Line Mapper array has been on sky since May 2022, with the majority of this time dedicated to debugging the updated observing software after switching from Intel Compute Sticks to Raspberry Pi computers, and taking calibration frames. The collection of commissioning data is currently in progress. In a future paper we will discuss the reduction in sky noise and signal to noise improvements going from 3nm to 0.8nm filters, as well as the the data reduction pipeline utilising all calibration filters. After analysing the commissioning data, we will finalize the filter bandpasses used on the 120 lens array and evaluate if any further system changes need to be made. If no major changes need to be made, then at a rate of one mount every three months, we will finish construction on the full 120-lens Dragonfly Spectral Line Mapper by mid 2023. 



\section{Summary}

The construction of the Dragonfly Spectral Line Mapper is currently in progress. This telescope is based on the Dragonfly Telephoto Array, with additional instrumentation that enables it to obtain ultra-narrow bandpass imaging of the low surface brightness universe. Building upon the success of the 3-lens pathfinder, we implement a number of upgrades in the first 10 lenses of the Dragonfly Spectral Line Mapper. We changed the H$\alpha$ filter bandpass from a 3nm to 0.8nm, and added \textsc{[Oiii]} filters as well as a suite of calibration filters. The H$\alpha$ and \textsc{[Oiii]} filters were tested in the laboratory to characterize their bandpass as a function of tilt angle. We improved the Filter-Tilter tilt accuracy to 0.2$^{\circ}$ by upgrading actuator system, and characterized the dark current of the Aluma 694 cameras used on the array. After obtaining sufficient commissioning data with the 10-lens array that is currently on-sky, we will continue construction on the full 120-lens array, and are on track to completion by mid 2023.

\appendix    


\acknowledgments 

We are very grateful to the staff at New Mexico Skies Observatories. Their support has been crucial to this project. We are thankful for contributions from the Dunlap Institute (funded through an endowment established by the David Dunlap family and the University of Toronto) which made this research possible. This research made use of Astropy,\footnote{http://www.astropy.org} a community-developed core Python package for Astronomy \cite{astropy:2013, astropy:2018}. We acknowledge the support of the Natural Sciences and Engineering Research Council of Canada (NSERC). Nous remercions le Conseil de recherches en sciences naturelles et en génie du Canada (CRSNG) de son soutien. We thank the the Canada Foundation for Innovation (CFI) for their support.


\bibliography{report} 
\bibliographystyle{spiebib} 

\end{document}